# Ni-PZT-Ni Trilayered Magnetoelectric composites Synthesized by Electro-deposition

Pan D A, Bai Y, Chu W Y and Qiao L J

Environmental Fracture Laboratory of Education, Corrosion and Protection Center, University of Science and Technology Beijing, 100083, P. R. China

E-mai: lqiao@ustb.edu.cn

PACS: *71.20.Nr*, *71.20.Be*

**Abstract**: We report the high strength of magnetoelectric (ME) coupling of trilayered composites prepared by electro-deposition. The ME coupling of Ni-lead zirconate titanate (PZT)-Ni trilayered structure was measured ranged from1 kHz to 120 kHz. The trilayered composites exhibit high magnetoelectric voltage coefficient because of good bonding between piezoelectric and magnetostrictive layers. The maximum magnetoelectric voltage coefficient can be up to 33 V/cm Oe at the electromechanical resonance frequency. This magnetoelectric effect shows promising application in transducers for magnetoelectric energy conversion.

## 1. Introduction

Multiferroic materials have drawn increasing interest due to their multi-functionality, which provides significant potentials for applications in the next-generation multifunctional devices [1]. In the multiferroic materials, the coupling interaction between multiferroic orders could produce some new effects, such as magnetoelectric (ME) or magnetodielectric effect [2]. The ME response, characterized by the appearance of an electric polarization upon applying a magnetic field and/or the appearance of magnetization upon applying an electric field, has been observed as an intrinsic effect in some single phase materials (e.g., $HoMnO_3$ at low temperature and in high magnetic field) [3]. Alternatively, multiferroic composites made by combination of ferromagnetic and ferroelectric substances such as piezoelectric ceramics [e.g., $BaTiO_3$ and lead zirconate titanate (PZT)] and ferrites were found to exhibit large room-temperature extrinsic ME effects recently [4-8], which has been known as a product property [9], i.e., a new property of such composites that either individual component phase does not exhibit. This ME effect can be defined as a coupling of magnetic-mechanical-dielectric behavior. That is, when a magnetic field is applied to the composites, the ferromagnetic phase changes the shape magnetostrictivity, and then the strain is passed along to the piezoelectric phase, resulting in an electric polarization [10].

In the past several years, various layered ME composites were widely investigated. Laletin et al reported the ME interactions in layered transition metal (TM)/PZT samples synthesized by bonding thin disks of PZT and Fe, Co or Ni [11]. To achieve better magnetoelectric properties, giant magnetostrictive material $Tb_{1-x}Dy_xFe_{2-y}$ (Terfenol-D) were used to combine with piezoelectric materials, such as PZT and polyvinylidene fluoride (PVDF), in a laminate structure [12-18]. These ME voltage coefficient were around 5 V/cm Oe.

Electro-deposition is widely used in preparation many composite function materials, composed of metal and alloys with good adhesion. Electro-deposition has outstanding other merits that it can be carried out on complex shape, and the thickness of plate is easy to be controlled.

## 2. Experimental details

PZT were first sliced in 10×20×0.25 mm, materized use sinter sliver paint at high temperature, and jointed electrode at both sides. PZT were polarized at 425 K in an electric field of 30-50 kV/cm perpendicular to the sample plane. Then, PZT was bathed in nickel aminosulfonate plating solution, and 5 $A/dm^2$ cathodic current density was applied to electroplate Ni on both sides of PZT. The compositions of plating solution and processing parameters are listed in Table I. Nickel aminosulfonate plating solution was used because of its advantages such as the solution stability, rapid plating speed and small internal stress. Before testing ME voltage coefficient, the layered compositions were dried for 6 days at 100 $^{o}$C in an oven.

Table I. Components and process parameters of the nickel electro-deposition

| | |
|---|---|
| Nickel aminosulfonate(g/l) | 600 |
| Nickelous chloride(g/l) | 20 |
| Boric acid(g/l) | 20 |
| Sodium lauryl sulfate(g/l) | 0.1 |
| PH | 4 |
| Temperature($^{o}$C) | 60 |
| Cathodic current density (A/dm$^2$) | 5 |

For ME measurement, the samples were subjected to a bias magnetic field $H$ and an AC field $\delta H$ (20 Hz-100 kHz). Since AC magnetic field $\delta H$ was generated by a Helmholtz coil, the amplitude of AC magnetic field $\delta H$ = 22 Oe when the amplitude of AC current is equal to 1A through the coil. The generated voltage $\delta V$ across the sample was amplified and measured with an oscilloscope. The ME voltage coefficient was estimated based on $\alpha_E = \delta V / (t_{\mathrm{PZT}} \cdot \delta H)$ , where $t_{\mathrm{PZT}}$ is the thickness of PZT. The measurements were carried out for two different field orientations. The transverse coefficient $\alpha_{\mathrm{E},31}$ was measured for $H$ and $\delta H$ parallel to the length of samples (direction-1) and perpendicular to $\delta E$ (direction-3). The longitudinal coefficient $\alpha_{\mathrm{E},33}$, measured for all the fields perpendicular to the sample plane (direction-3).

## 3. Results and discussion

Firstly, the dependence of $\alpha_{\mathrm{E}}$ on $H$ was measured at 1 kHz for transverse (in-plane) and longitude (out-of-plane) magnetic fields, respectively. (Fig. 1) With the rise of $H$, $\alpha_{\mathrm{E},31}$ increases first, reaches a maximum at $H_{\mathrm{m}}$=0.16 kOe, then decreases rapidly. In contrast, with the rise of $H$, $\alpha_{\mathrm{E},33}$ first increases to a maximum at $H_{\mathrm{m}}$=4.5 kOe, and then decreases slowly. The magnitude and the field dependence of $\alpha_E$ are related to variation of the demagnetic effect [19]. The ME coefficients are directly proportional to $q \sim \delta\lambda/\delta H$, where $\delta\lambda$ is the magnetostriction, and the $H$-dependence tracks the slope of $\lambda$ vs $H$. Saturation of $\lambda$ at high field leads to $\alpha_E$=0.

Secondly, $\alpha_E$ was measured at the bias field of $H_m$ as frequency of AC magnetic field *(f)* varied from 1 kHz to 120 kHz. (Fig. 2) Typical $\alpha_E$ vs $f$ profile for transverse fields is shown in Fig. 2(a). Both for $\alpha_{E,31}$ and $\alpha_{E,33}$, there is a sharp peak at about 88 kHz. However, the maximum of $\alpha_{E,33}$, about 1.08 V/cm Oe, is an order of magnitude smaller than that of $\alpha_{E,31}$, 33 V/cm Oe.

Figure 3 shows the frequency dependence of dielectric constant and dielectric loss of the laminated Ni-PZT-Ni composite. There is a resonance peak at about 89.9 kHz which is associated with the electromechanical resonance (EMR) [20]. The consistency of the frequency of the peak of dielectric constant and $\alpha_E$ indicates that the high ME effect is associated with the EMR.

Figure 2 inset shows frequency dependence of $\alpha_{E,31}$ around the frequency of EMR for the samples with different Ni thickness, $t_{Ni}$. The frequency of EMR shifts towards high frequency with the rise of Ni thickness, because it is directly proportional to the thickness of piezoelectric and piezomagnetic layers [21]. Figure 3 insert also shows frequency dependence of dielectric constant is consistent with that of EMR for the samples with different Ni thickness. The resonance frequency of magnetoelectric voltage coefficient in Fig. 2 inset is corresponding well that of dielectric constant in Fig. 3 inset for the samples with the same Ni thickness.

Figure 4 plots the dependence of ME voltage coefficient on $t_{Ni}/(t_{Ni} + t_{PZT})$ for Ni-PZT-Ni trilayered at the frequency of EMR. Both $\alpha_{E,31}$ and $\alpha_{E,33}$ increase with the ratio of $t_{Ni}/(t_{Ni} + t_{PZT})$. When the total thickness of Ni is about 0.4mm, $\alpha_{E,31}$ is up to 33 V/cm Oe, and $\alpha_{E,33}$ is up to 1.08 V/cm Oe. These results can be well reproduced and have good agreement with theoretical expression for layered ME composites [22]:

$$\alpha_{E,31} = \frac{-k(q_{21}+q_{11})d_{31}t_{\mathrm{Ni}}t_{\mathrm{PZT}}}{(s_{11}^{\mathrm{Ni}}+s_{12}^{\mathrm{Ni}})\varepsilon_{33}kt_{\mathrm{PZT}}+(s_{11}^{\mathrm{PZT}}+s_{12}^{\mathrm{PZT}})\varepsilon_{33}t_{\mathrm{PZT}}-2(d_{31})^2kt_{\mathrm{Ni}}}$$

$$\alpha_{E,33} = \frac{-2kq_{31}d_{31}t_{\mathrm{Ni}}t_{\mathrm{PZT}}}{(s_{11}^{\mathrm{Ni}}+s_{12}^{\mathrm{Ni}})\varepsilon_{33}kt_{\mathrm{PZT}}+(s_{11}^{\mathrm{PZT}}+s_{12}^{\mathrm{PZT}})\varepsilon_{33}t_{\mathrm{Ni}}-2(d_{31})^2kt_{\mathrm{Ni}}} \quad (1)$$

where $k$ is interface coupling parameter, $d_{31}$ is the piezoelectric constant of PZT, $t_{\mathrm{Ni}}$ is the thickness of Ni, $t_{\mathrm{PZT}}$ is the thickness of PZT, $S$ is the compliance coefficients of Ni or PZT, and $\varepsilon_{33}$ is the dielectric constant of PZT.

Since the thickness of PZT is constant in our experiment, the ME voltage coefficient is directly proportional to the thickness of Ni. Based on equation (1), $\alpha_{\mathrm{E},31}$ should increases with increasing $t_{\mathrm{Ni}}$ because of $d(\alpha_{\mathrm{E},31})/d(t_{\mathrm{Ni}})$ >0. Nan et al theoretically reported that $\alpha_{\mathrm{E}}$ increases with $t_{\mathrm{Ni}}$ monotonously [23]. Our experimental result is in good agreement with this theoretical prediction.

For the magnetoelectric laminate structure, the interfacial binding between magnetostrictive layer and piezoelectric layer is important to the magnetoelectric coefficient. Liu et al theoretical reported the influence of interfacial binder layer's thickness and shear modulus on the magnetoelectric effect [24]. While interfacial binder layer's thickness rises or its shear modulus reduces, the ME responses will decrease rapidly. It is worthwhile to note that although the magnetostriction of Ni is two orders of magnitude smaller than that of Terfenol–D, the coefficient $\alpha_{\mathrm{E}}$ of the laminated Ni-PZT-Ni composite synthesized by electro-deposition is comparative to that of Terfenol–D/PZT/PVDF bulk samples [25, 26]. It is because the plastic interfacial layer of PVDF in the Terfenol–D/PZT/PVDF system was replaced by Ag metal interfacial layer in the electro-deposited Ni-PZT-Ni system. The shear modulus of Ag is much higher than that of PVDF, and the Ag layer is thinner than PVDF bonder layer. Hence, better interfacial coupling between PZT and Ni layers will supply the gap of Ni's small magnetostriction. It is promising to enhance the ME coefficient more by improving electro-deposition technology, such as electro-depositing metal with higher magnetostriction coefficient.

## 4. Summary

In summary, this article presents first report on ME interaction of Ni-PZT-Ni trilayered composites synthesized by electro-deposition. At the frequency of EMR, the coefficient $\alpha_E$ have a peak maximum, $\alpha_{E,31}$ up to 33 V/cm Oe when the thickness of Ni is about 0.4 mm. Tight bonding between PZT and Ni layers makes these samples exhibit large ME voltage coefficient among bulk magnetoelectric composites. Electro-deposition provide an effective method to enhance ME coefficient of magnetoelectric composites remarkably. Moreover, the electro-deposition method makes the preparation of ME composites with complex shape easily, and can control the structural parameters effectively. It will promote a rapid development of magnetoelectric composites' applications, such as various magnetoelectric coupling devices.

**ACKNOWLEDGMENTS**

The authors acknowledge the group of C.-W. Nan(State Key Laboratory of New Ceramics and Fine Processing, Department of Materials Science and Engineering, Tsinghua University) for the help of ME voltage coefficient testing. This project was supported by program for Changjiang Scholars, Innovative Research Team in University (IRT 0509) and the National Natural Science Foundation of China under Grant No. 50572006.

## Figure Captions Page

FIG. 1. Magnetoelectric voltage coefficient $\alpha_{E,31}$ and $\alpha_{E,33}$ at room temperature for Ni-PZT-Ni trilayered composites with total thickness of Ni about 0.4 mm.

FIG. 2. Frequency dependence of $\alpha_{E,31}$ (a) and $\alpha_{E,33}$ (b) for the Ni-PZT-Ni trilayered composites with total thickness of Ni about 0.4mm at $H_m$ corresponding to maximum ME coupling (see FIG. 1.). The inset shows ME voltage coefficient around EMR frequency for the samples with different Ni thickness.

FIG. 3. Frequency dependence of dielectric constant and dielectric loss for the Ni-PZT-Ni trilayered composites with total thickness of Ni about 0.4 mm. The inset shows dielectric constant around EMR frequency for the samples with different Ni thickness.

FIG. 4. $t_{\mathrm{Ni}}/(t_{\mathrm{N}i}+t_{\mathrm{PZT}})$ dependence of ME voltage coefficient for Ni-PZT-Ni trilayered composites at the frequency of EMR.

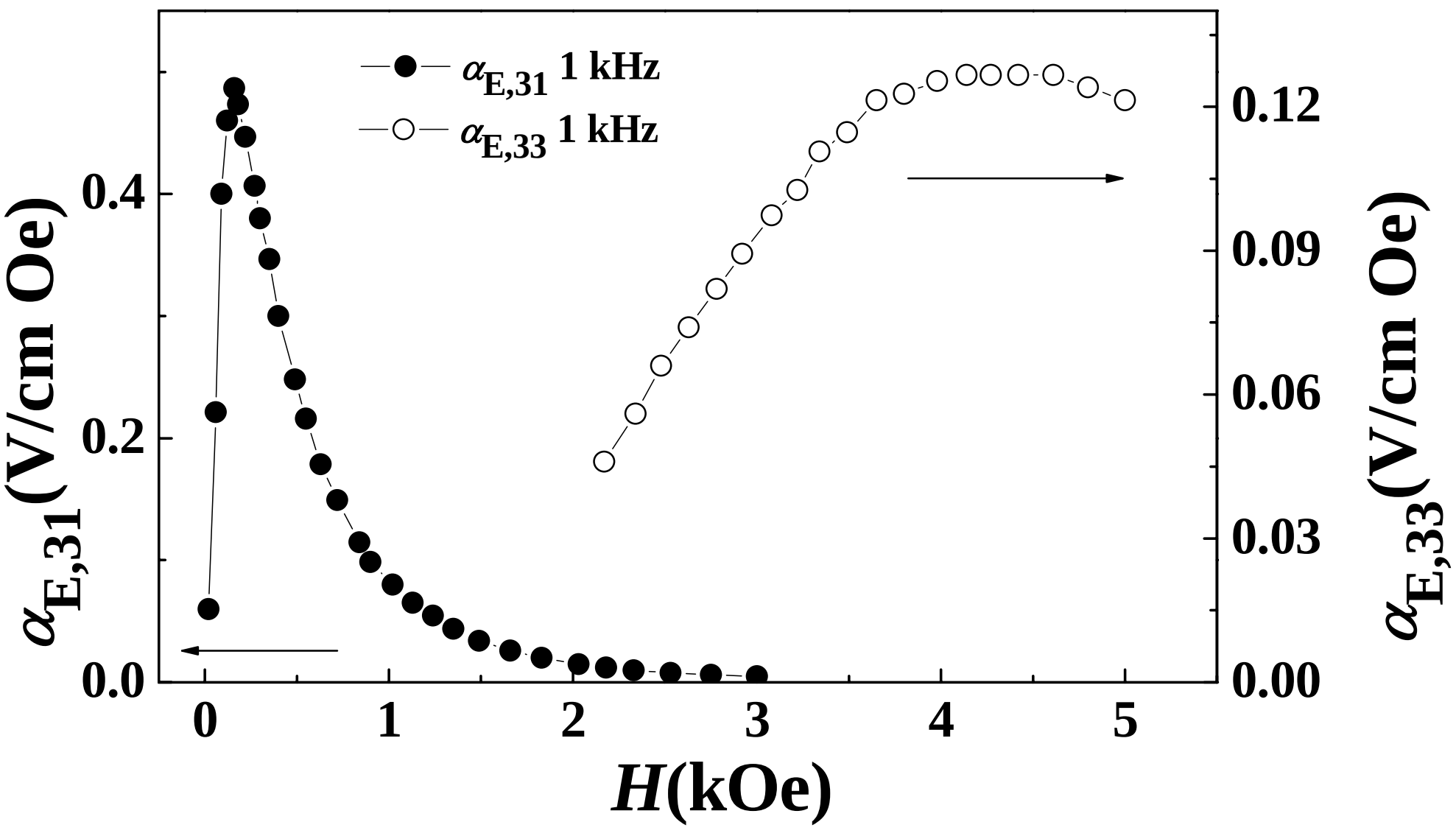

FIG. 1. Magnetoelectric voltage coefficient $\alpha_{E,31}$ and $\alpha_{E,33}$ at room temperature for Ni-PZT-Ni trilayers with total thickness of Ni about 0.4 mm.

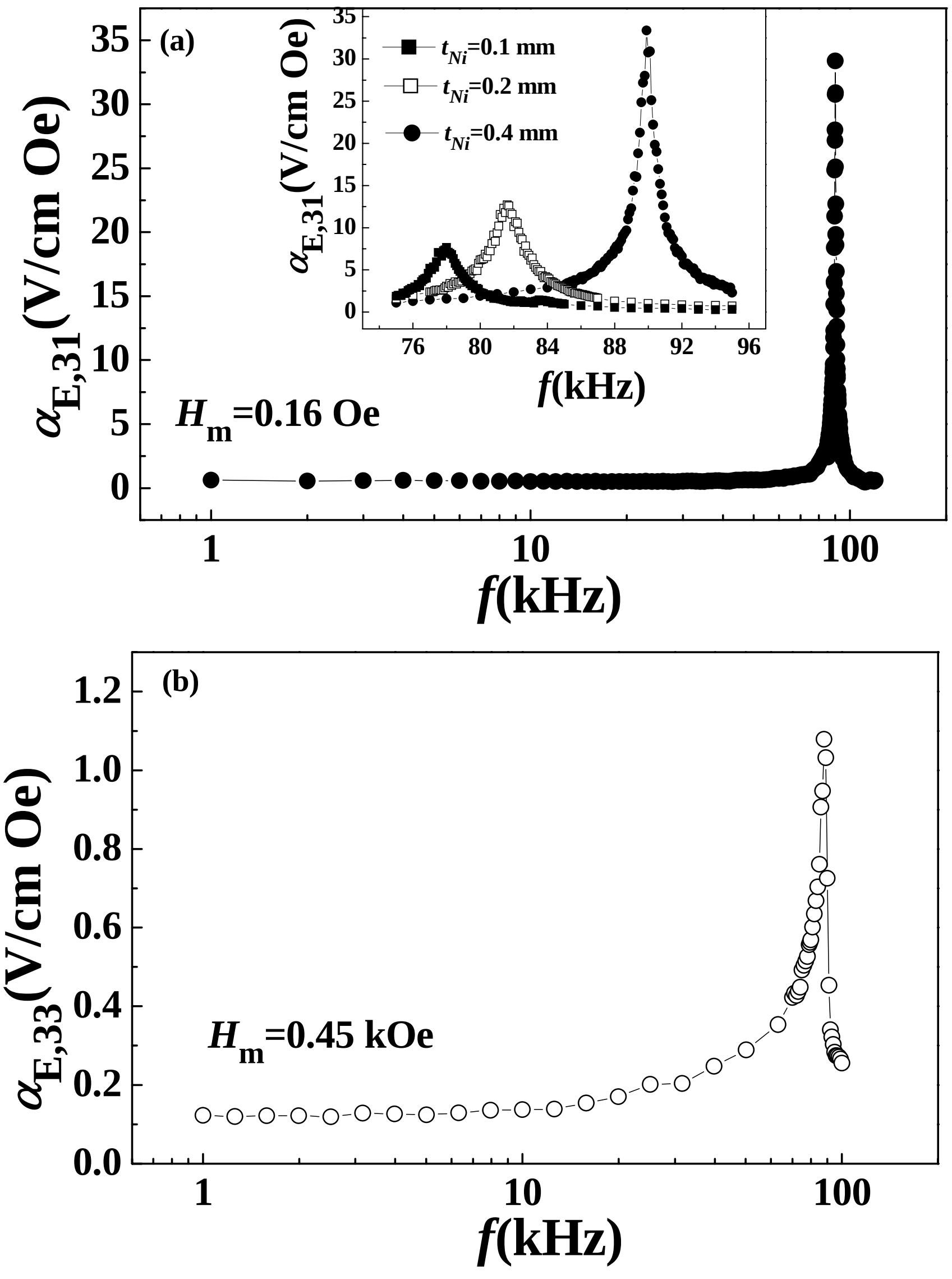

FIG. 2. Frequency dependence of $\alpha_{E,31}$ (a) and $\alpha_{E,33}$ (b) for the Ni-PZT-Ni with total thickness of Ni about 0.4mm at $H_m$ corresponding to maximum ME coupling (see FIG. 1.). The inset shows ME voltage coefficient around EMR frequency for the samples with different Ni thickness.

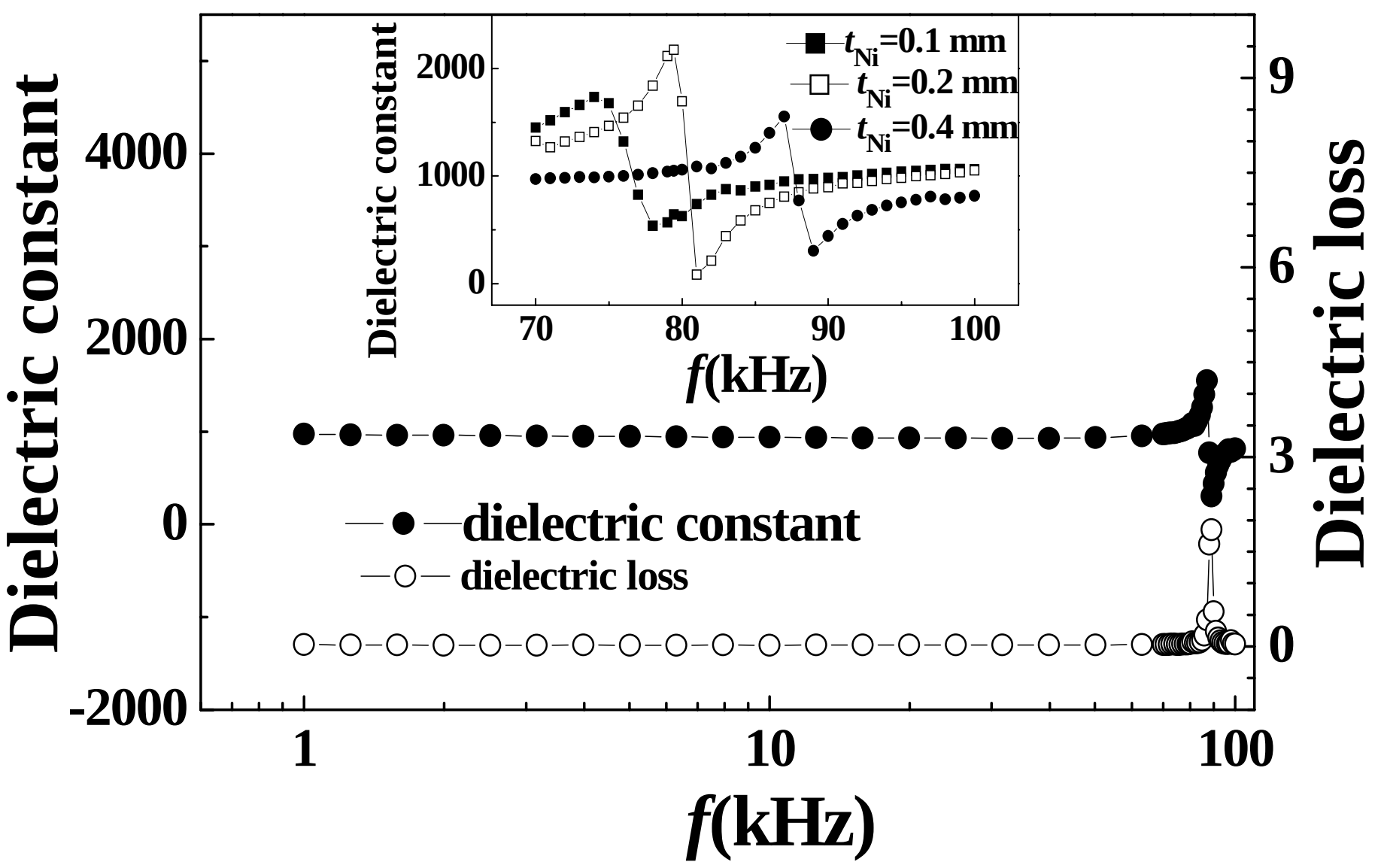

FIG. 3. Frequency dependence of dielectric constant and dielectric loss for the Ni-PZT-Ni composites with total thickness of Ni about 0.4 mm. The inset shows dielectric constant around EMR frequency for the samples with different Ni thickness.

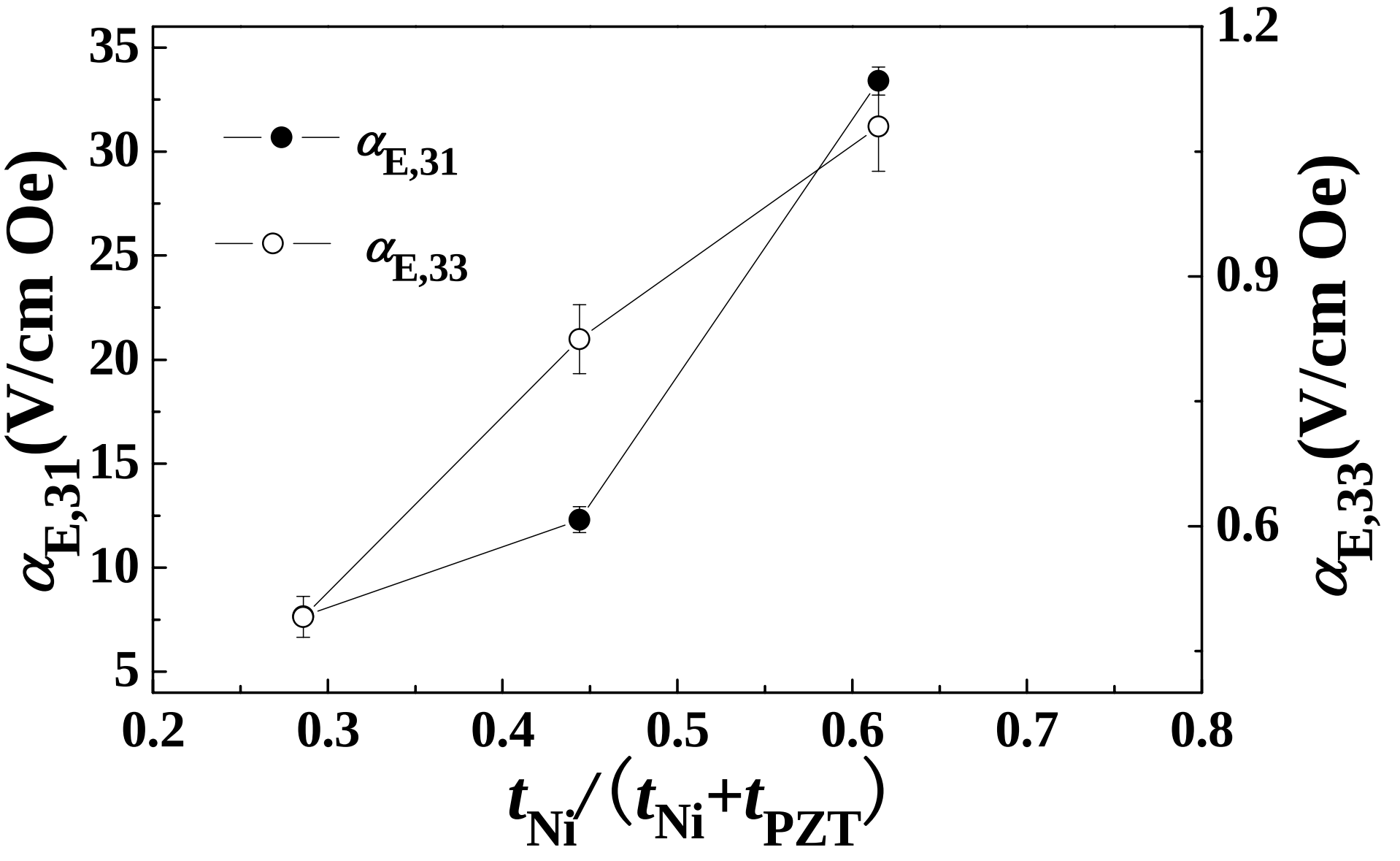

FIG. 4. $t_{\mathrm{Ni}}/(t_{\mathrm{N}i}+t_{\mathrm{PZT}})$ dependence of ME voltage coefficient for Ni-PZT-Ni trilayers at the frequency of EMR.